\documentstyle[preprint,aps]{revtex}
\tightenlines
\font\eu=eufm10
\begin{document}
\draft
\font\eu=eufm10

\title{ON THE LENZ-ISING-ONSAGER PROBLEM IN AN EXTERNAL MAGNETIC FIELD}

\author{Martin S. Kochma\'nski \\}
\address{Institute of Physics, 
University of Rzesz\'ow\\
T.Rejtana 16 A , 35--310 Rzesz\'ow, Poland\\
e-mail: mkochma@atena.univ.rzeszow.pl}
\maketitle
\begin{abstract}
The Lenz-Ising-Onsager (LIO) problem in an external magnetic field in the 
second quantization representation is the subject of consideration of the 
paper. It is shown that the operator $V_h$ in the second quantization 
representation corresponding to Ising spins interaction with the external 
magnetic field $H$ can be represented in terms of single-subscript creation 
and anihilation Fermi operators in such a form that the operator $V_h$ 
commutes with the operator  $\hat{P}\equiv(-1)^{\hat{S}}$, where  $\hat{S}=
\sum_m\beta^{\dag}_m\beta_m$  is the operator of a total number of Fermions. 
The possible consequences of such representation with it's relation to the 
LIO is discussed. In particular, the constructive proof of the Lee-Yang 
theorem on the absence of phase transition for Ising model in nonzero 
magnetic field $(\Re h\neq 0)$ is demonstrated.
\end{abstract}

\widetext
\section{Introduction}
As is known, up to now there is no solution for the $2D$ Ising model (IM) in 
an external magnetic field $H$ (Lenz-Ising-Onsager problem - LIO) still. The 
algebraic, topological and graph-theoretical difficulties of taking into 
account the external magnetic field in LIO problem are well known and widely 
discussed $\cite{onsager44,onkauf49,sml64,feyn72,ziman79,war92,war94,grim97,koch97}$. 
As is known $\cite{sml64}$, the statistical sum for the $2D$ Ising model in 
external field $(H)$ in the representation of second quantization can be 
written in the form:
\begin{equation}
Z_{2D}(h)=Tr(V)^{N}=Tr(V_{1}V_{2}V_{h})^{N},
\end{equation}
where the operators $\it V_{i}$, expressed in terms of the Fermi creation and 
annihilation operators ($c^{\dag}_{m}, c_{m}$), are of the form: 
\begin{equation}
\begin{array}{c}
V_{1}=(2{\sinh}2K_{1})^{M/2}\exp{\left[-2K_{1}^{*}\sum_{m=1}^{M}(c_{m}^{\dag}
c_{m}-1/2)\right]},\\
V_{2}=\exp\left\{K_{2}\left[\sum_{m=1}^{M-1}(c_{m}^{\dag}-c_{m})(c_{m+1}^
{\dag}+c_{m+1}) -(-1)^{\hat{M}}(c_{M}^{\dag}-c_{M})(c_{1}^{\dag}+c_{1})\right]
\right\},\\
V_{h}=\exp\left\{h\sum_{m=1}^{M}\exp\left[i\pi\sum_{p=1}^{m-1}
c_{p}^{\dag}c_{p}\right](c_{m}^{\dag}+c_{m})\right\},
\end{array}
\end{equation}
where $K_{j}=\beta J_j;\;(j=1,2,)$ and $h=\beta H$ ($\beta =1/k_BT$ - inverse 
temperature, $H$ - vnexnee magnitnoe pole), and $\hat{M}=\sum_{1}^{M}c_{m}^{\dag}c_{m}$ 
is the operator of the total number of particles and $K_{1}^{*}$ and  
$K_{1}$ are connected by the following formulae:
\begin{equation}
\tanh(K_{1})=\exp(-2K_{1}^{*}), \;\;\; or \;\;\; \sinh2K_{1}\sinh2K_{1}^{*}=1.
\end{equation}
One can see that the operator $V_h$ in the second quantization representation, 
that describes interaction of the spins with external magnetic field, has 
rather complicated structure. It is easy to see that this operator does not 
commute with the operator $\hat{P}\equiv(-1)^{\hat{M}}$. As a result the 
operator $V_2$ has also not a very tractable form, i.e. it has not the needed 
translational symmetry (2). More exactly, although  the operators $V_1$ 
and $V_2$ commute with the operator $\hat{P}$, the operator $V$ (1) does 
not commute with the operator $\hat{P}$, i.e. $[\hat{P},V]_{-}\neq{0}$, 
because $[\hat{P},V_h]_{-}\neq{0}$. Therefore, we can not divide all states 
of the operator $V=V_{1}V_{2}V_h$ into eigenstates of the operator $\hat{P}$ 
with eigenvalues $\lambda=\pm 1$, and this leads to nonconservation of the 
states with even and odd numbers of fermions (for details see $\cite{sml64}$). 
Namely this is the fundamental reason which stops solving the problem under 
consideration within this formalism. 

Coming back to the difficulties mentioned above which are connected with the 
operator $V_h$, (2), it is now clear that to overcome the troubles within 
the approach $\cite{sml64}$, one should find an appropriate method of 
substituting the operator $V_h$ (2) with another one which would be 
equivalent to the former in the sense of correct counting of the interaction 
of external magnetic field with the spins of the system. Namely, as it could 
be easily seen, the only contribution to $Z_{2D}$ (1) from the operator $V_h$ 
comes, in the representation of second quantization, from the "even" part 
with respect to operators $c^{\dag}_{m},c_{m}$ of the operator $V_h$.  

It was shown in the author's paper $\cite{koch97}$  how this difficulty can 
be overcome by transition to the space of higher dimensionality and supposing 
then one of the interaction constants to be equal to zero. Then all the 
operators $V_j$ in the second quantization representation are expressed in 
terms of two-subscript creation and anihilation Fermi operators in the 
finite-dimensional Fock space $2^{NM}$  in the particle number representation. 
These operators $V_j$ commute with the operator $\hat{P}=(-1)^{\hat{S}}$, 
where $\hat{S}=\sum_{nm}\alpha^{\dag}_{nm}\alpha_{nm}$ is the operator of 
total number of fermions. Here we discuss briefly how the analogous result 
could be achieved in the space of single-subscript Fermi operators, that is, 
in the finite-dimensional Fock space  $2^M$. In particular, I'm intend to 
sketch briefly that the partition function for the $2D$ Ising model in an 
external magnetic field can be expressed in the form:
\begin{equation}
\begin{array}{c}
Z_{2D}(h)=Tr(V)^{N}=Tr(V_{1}V_{2}V_{h})^{N},\\
V_{1}=(2{\sinh}2K_{1})^{M/2}\exp{\left[-2K_{1}^{*}\sum_{m=1}^{M}
(\beta_{m}^{\dag}\beta_{m}-1/2)\right]},\\
V_{2}=\exp\left\{K_{2}\left[\sum_{m=1}^{M}(\beta_{m}^{\dag}-\beta_{m})
(\beta_{m+1}^{\dag}+\beta_{m+1})\right]\right\},
\end{array}
\end{equation}
\begin{equation}
V_{h}=(\cosh h)^M\prod_{m=1}^M\prod_{k=1}^{M-m}\left[1+\tanh^2h
(\beta^{\dag}_m-\beta_m)(\beta^{\dag}_{m+k}+\beta_{m+k})(-1)^
{\sum_{p=m+1}^{m+k-1}\beta^{\dag}_p\beta_p}\right].
\end{equation}
It is quite obvious that the operator  $V_h$ (5) commutes with the operator
$\hat{P}=(-1)^{\hat{S}}$, where $\hat{S}=\sum_{m=1}^M\beta^{\dag}_p\beta_p$ 
is the operator of total number of fermions, and hence, the operator $V_2$ 
(4) also can be expressed in that form, just as it is done here, that is 
without term $\sim (-1)^{\hat{S}}$ unlike the representation (2). 

\section{THE PARTITION FUNCTION }

The main idea of representation (4)-(5) is to consider the $2D$ Ising model in 
an external magnetic field in terms of second quantization representation, in 
in the space of two-subscript creation and anihilation Fermi operators and 
after that to factorize the corresponding operators $T_{1,2}$ (see below). 
Namely, it was shown in the author's  $\cite{koch97}$ that the partition 
function of $2D$ IM in an external magnetic field is of the form:
\begin{equation}
Z_{2D}(h)=(2\cosh h)^{NM}\left(\prod_{0<{q,p}<\pi}A^2_1(q)\right)
\left(\prod_{0<{q,p}<\pi}A^2_2(p,h)\right)\langle 0|T_2(h)T_1|0\rangle,
\end{equation} 
where
\begin{equation}
\begin{array}{c}
T_1=\exp\left[\sum^{N}_{n=1}\sum^{N-n}_{l=1}\sum^{M}_{m=1}a(l)
\beta^{\dag}_{nm}\beta^{\dag}_{n+l,m}\right], \;\;
a(l)=\frac{1}{N}\sum_{0<{q}<\pi}2B_1(q)\sin(lq); \\
T_2(h)=\exp\left[\sum^{N}_{n=1}\sum^{M}_{m=1}\sum^{M-m}_{k=1}b(k)
\alpha_{n,m+k}\alpha_{nm}\right],\;\;
b(k)=\frac{1}{M}\sum_{0<{p}<\pi}2B_2(p)\sin(kp),\\
\end{array}
\end{equation}
and
\begin{equation}
\begin{array}{c}
A_2(p,h)=\cosh 2K_2-\sinh 2K_2\cos p+\alpha(h)\sinh 2K_2\sin p,\\
A_1(q)=\cosh 2K_1-\sinh 2K_1\cos q,\;\;\;\;\alpha(h)=\tanh^2h\frac{1+\cos p}
{\sin p},\\
B_2(p)=\frac{\alpha(h)[\cosh 2K_2+\sinh 2K_2\cos p]+\sinh 2K_2\sin p}
{A_2(p,h)}, \;\;\;B_1(q)=\frac{\sinh 2K_1\sin q}{A_1(q)}.
\end{array}
\end{equation}
Here $(\alpha^{\dag}_{nm},\alpha_{nm})$ and $(\beta^{\dag}_{nm},\beta_{nm})$ 
are the two-subscript Fermi operators, $|0\rangle$ is the fermionic vacuum 
function in the finite-dimensional Fock space of $2^{NM}$ dimensions in the 
occupation number representation. Fermi operators $\alpha$- and $\beta$-  
are related to each other by canonical unitary transformations 
$(\alpha^{\dag}_{nm}\alpha_{nm}=\beta^{\dag}_{nm}\beta_{nm})$:
\begin{equation}
\begin{array}{c}
\alpha^{\dag}_{nm}(\alpha_{nm})=\exp{(i\pi\phi_{nm})\beta^{\dag}_{nm}(\beta_{nm})},
\nonumber\\
\phi_{nm}=\left[\sum^{N}_{k=n+1}\sum^{m-1}_{p=1}+\sum^{n-1}_{k=1}
\sum^M_{p=m+1}\right]\alpha^{\dag}_{kp}\alpha_{kp}=[\ldots]\beta^{\dag}
_{kp}\beta_{kp}.
\end{array}
\end{equation}
In order to transform the representation (6)-(8) into (4)-(5), one should make 
the following: to rewrite the operator {$T_1$ (7) in terms of Pauli operators 
($\tau^{\pm}_{nm}$), to apply the well-known representation 
\begin{eqnarray*}
\exp(\hat{C}^2) = \frac{1}{\sqrt{\pi}}\int_{-\infty}^{\infty}\exp[-\xi^2+
2\hat{C}\xi]d\xi,
\end{eqnarray*}
(which is correct for the totally bounded operators), to turn Pauli operators 
($\tau^{\pm}_{nm}$) into to the Fermi ones $(\alpha^{\dag}_{nm},
\alpha_{nm})$, then to "drag" the emerging phase factors through $|0\rangle$, 
$\cite{koch97}$, to introduce Ising-type variables $(\mu_{nm}=\pm 1)$, 
to apply for the second time transfer-matrix method and finally to use the 
one-dimensional Jordan-Wigner transformation. Omitting the great number of 
of intermediate calculations, for the partition function (6) we have: 
\begin{equation}
Z_{2D}(h)=(\cosh h)^{NM}\cdot Tr(V_{1}V_{2}V^*_{h})^{N},
\end{equation}
where operators $V_{1,2}$ is defined by formulae (4) and trace is calculated 
by means of $\beta$-operators, where operator $V_h^*$ is of the form:
\begin{equation}
V^*_h=\langle 0|e^{\tanh^2h\sum_{m=1}^M\sum_{k=1}^{M-m}\alpha_{m+k}\alpha_m}
\prod_{m=1}^M\left[1+(\alpha^{\dag}_m-\alpha_m)(\alpha^{\dag}_{m+1}+\alpha_{m+1})
\cdot (\beta^{\dag}_m-\beta_m)(\beta^{\dag}_{m+1}+\beta_{m+1})\right]|0\rangle_
{\alpha}.
\end{equation}
Here the single-subscript $\alpha$- and $\beta$- Fermi operators are absolutely 
independent ones (as they are introduced), commuting or anti-commuting with 
each other (the final result does not depend on the fact, as it should be 
expected) and the vacuum martix element is calculated by means of $\alpha$-
operators. Using Wick theorem, one can easily calculate the vacuum matrix 
element in (11) and obtain the final expression (5) for the operator $V_h$, 
taking into account the factor $(\cosh h)^{NM}$ at the  $Tr(...)$ 
in (10). One can say, using the figure of speech that the operator 
$\exp[\tanh^2h\sum_{m=1}^M\sum_{k=1}^{M-m}\alpha_{m+k}\alpha_m]$ within the 
brackets $\langle 0|...|0\rangle$ acts on the operator $\prod_m[...]$ just 
like the "Ockham razor" does, cutting off, due to Wick theorem 
$\cite{wick50}$, all the superfluous terms. 

From (4)-(5) one can derive some consequences, in particular, the constructive 
proof of well-known Lee-Yang theorem $\cite{lee-yang52,glimm81}$ on the 
absence of phase transition for the Ising model in nonzero magnetic field 
$\Re h\neq 0$. Indeed, it can be easily shown from (4)-(5) that at $h=0$ we have 
Onsager solution and supposing $K_1$ to be zero ($K_1=0$), we have next 
representation for the partition function: 
\begin{eqnarray*}
\begin{array}{c}
Z_{1D}=(2\cosh h)^{M}\langle0\mid(V_{2}V_{h}^{*})\mid0\rangle,\\
V_{h}^{*}=\exp\left[\tanh^2h\sum_{m=1}^{M}\sum_{p=1}^{M-m}
\beta_{m}^{\dag}\beta_{m+p}^{\dag}\right], 
\end{array}
\end{eqnarray*}
from which we can derive exactly the classic Ising result $\cite{koch97}$. 

But it is not trivial to have by means of (4)-(5) Ising solution supposing $K_2$ 
to be zero $(K_2=0)$. For this case operator $V_2(K_2=0)=\hat{I}$ and the 
problem is reduced to the calculation of trace with the operator $(V_1V_h)^N$, 
which is not easy. However. as it follows from (4)-(5) by it's construction, 
we have here the Ising solution also. The last one allows to prove 
constructively the Lee-Yang theorem for Ising model 
$\cite{lee-yang52,glimm81}$. 

First of all, it is easy to demonstrate that at small but finite external 
magnetic field $(h\sim\varepsilon\ll 1)$, the expression for operator $V_h$, 
(5) can be represented as: 
\begin{equation}
V_h\cong\exp\left[h^2\sum_{m=1}^M\sum_{k=1}^{M-m}
(\beta^{\dag}_m-\beta_m)(\beta^{\dag}_{m+k}+\beta_{m+k})(-1)^
{\sum_{p=m+1}^{m+k-1}\beta^{\dag}_p\beta_p}\right],
\end{equation}
with the accuracy up to $\sim\varepsilon^4$. It is obvious that the next 
Hamiltonian: 
\begin{equation}
{\cal H}=-\sum_{n=1,m=1}^{N,M}\Big(J_1\sigma_{nm}\sigma_{n+1,m} + 
J_2\sigma_{nm}\sigma_{n,m+1}\Big)-\frac{H^2}{k_BT}
\sum_{n=1,m=1}^{N,M}\sum_{k=1}^{M-m}\sigma_{nm}\sigma_{n,m+k} , 
\end{equation}
corresponds to the partition function (4) with the operator $V_h$ (12), with 
the accuracy up to unessential constant $(\sim H^2/2k_BT)$, where $T$ denotes 
temperature and $k_B$ the Boltzmann constant and $J_{1,2}$ are the interaction 
constants. Although the 
model (13) looks like asymmetric with respect to the second subscript $m$, 
actually this asymmetry does not influence on the final result. It is 
obvious, that at $J_1=0$ the Hamiltonian (13) describes $1D$ Ising model at 
small magnetic field. However, the situation is quite different, if $J_2=0$ 
and we have two-dimensional Ising model with the interaction constants $J_1$ 
and $J_2^*=H^2/k_BT$. For this model as it is stressed above, there is no 
phase transition at finite temperature because this modell actually describes 
$1D$ Ising model in small magnetic field $H$. It is obvious that the 
including an additional nearest neighbours interaction $J_2\neq 0$ 
corresponding to $2D$ Ising model in a small magnetic field 
$(H\sim\varepsilon\ll 1)$, does not lead to the phase transition also. The 
last one can be derived directly from the Hamiltonian (13). The same 
reasoning are also valid for the $3D$ Ising model in a small magnetic field; 
here it is also possible to introduce Hamiltonian, just like to (13) with 
the field-square term, using for this case the representation:
\begin{eqnarray*}
\begin{array}{c}
Z_3(h)=(2\cosh h)^{NMK}\langle 0|T_3T_2T_1T_h|0\rangle, \\
T_1=\exp\left[K_{1}\sum_{n,m,k=1}^{N,M,K}(\alpha^{\dag}_{nmk}-\alpha_{nmk})
(\alpha^{\dag}_{n+1,mk}+\alpha_{n+1,mk})\right] , \\
T_2=\exp\left[K_{2}\sum_{n,m,k=1}^{N,M,K}(\beta_{nmk}^{\dag}-\beta_{nmk})
(\beta^{\dag}_{n,m+1,k}+\beta_{n,m+1,k})\right] ,  \\
T_3=\exp\left[K_{3}\sum_{n,m,k=1}^{N,M,K}(\theta_{nmk}^{\dag}-\theta_{nmk})
(\theta^{\dag}_{nm,k+1}+\theta_{nm,k+1})\right] ,
\end{array}
\end{eqnarray*}
and ($\mu\equiv\tanh^2(h)$)
\begin{eqnarray*}
T_h=\exp\!\left\{\!\mu\left[\sum_{nmk}^{NMK}\sum^{N-n}_{s=1}
\alpha^{\dag}_{nmk}\alpha^{\dag}_{n+s,mk}+\sum_{nn'mk}^{NMK}\sum^{M-m}_{t=1}
\alpha^{\dag}_{nmk}\alpha^{\dag}_{n',m+t,k}+\sum_{nn'mm'k}^{NMK}\sum^{K-k}_
{l=1}\alpha^{\dag}_{nmk}\alpha^{\dag}_{n'm',k+l}\right]\!\right\},
\end{eqnarray*}
for the $3D$ Ising model in an external magnetic field $\cite{mkoch98}$, 
where $\alpha_{nmk}, ...$ are Fermi operators.

\section{Conclusions}

Despite the three representations (1)-(3), (4)-(5) and (6)-(8) for the 
partition function of the Ising model in an external magnetic field include 
"unpleasant" phase factors, the representation (4)-(5) has some advantages 
comparing to the (1)-(3) and might be to (6)-(8) ones. To the author's mind, 
the application of well-known approximate methods (analytical as well as 
numerical) to the Hamiltonian (13) can lead to the much better results than 
it took place in case of standard Hamiltonian of Ising model in an external 
magnetic field (with linear-field term). It is also possible that the 
application of the direct Onsager method $\cite{onsager44,onkauf49}$ to the 
operator $V\equiv V_1V_2V_h$ (4)-(5) could allow to find out the 
eigenfunctions and eigenvalues of that operator. 

I am grateful to I. Tralle for his assistance in preparation of the final 
form of this paper.

This paper was supported by the KBN grant $Nr$ {\bf 2 P03B 088 15}.

\end{document}